\documentstyle[aps,pre,epsf,twocolumn]{revtex}
\begin{document}
\def\break#1{\pagebreak \vspace*{#1}}
\def\bea{\begin{eqnarray}}
\def\eea{\end{eqnarray}}
\def\a{\alpha}
\def\D{\langle l \rangle}
\def\p{\partial} 
\draft
\title{ Equilibrium and dynamical properties of the
ANNNI chain at the multiphase point}  
\author{Abhishek Dhar, B. Sriram Shastry and Chandan Dasgupta}
\address{ Physics Department, Indian Institute of Science,
Bangalore 560012, India.\\}
\date{\today}
\maketitle
\widetext
\begin{abstract}
We study the equilibrium and dynamical properties of the
ANNNI (axial next-nearest-neighbor Ising) chain at the multiphase
point. An interesting 
property of the system is the macroscopic degeneracy of the ground
state leading to finite zero-temperature entropy. In our
equilibrium study we consider the effect of softening the spins. 
We show that the degeneracy of the ground
state is lifted and there is a qualitative change in the low
temperature behaviour of the system with a well defined low
temperature peak of the specific heat that carries the thermodynamic
``weight'' of the 
ground state entropy. In our study of the dynamical properties, the
stochastic Kawasaki dynamics is considered. 
The Fokker-Planck operator for the process corresponds to a quantum
spin Hamiltonian similar to the Heisenberg
ferromagnet but with constraints on allowed states. This leads to
a number of differences in its properties which are obtained through 
exact numerical diagonalization, simulations and by
obtaining various analytic bounds. 
\end{abstract}


\narrowtext

\section{Introduction}

The ANNNI chain is one of the simplest systems with competing
interactions. It is defined by the following Ising spin Hamiltonian:
\bea
H= \sum_{i=1}^L ( J_1  s_i s_{i+1} + J_2 s_i s_{i+2} ),~~~~s_i=\pm1.
\label{hard}
\eea
For $J_2 >0 $, the interactions are competing and one can have
different ground states depending on the relative
strengths of the  interactions. A specially interesting case is
the point $J_1=2 J_2$, the so-called multiphase point, where the ground
state is no longer unique. It can be shown that any spin
configuration, which does not have three 
consecutive spins of the same sign, is a ground state. For a chain
of length $L$, the number of ground states $\sim \mu^L$, where $\mu
=(\sqrt{5} + 1)/2$ is the golden mean. Thus there are an exponentially
large number of degenerate ground states and the system has finite
zero-temperature entropy per spin. The model has been
extensively studied both in one and higher
dimensions and is known to have a 
rich and interesting phase diagram \cite{solid}. In this paper we
consider some aspects of the 
equilibrium and dynamical behaviour of the ANNNI chain at 
the multiphase point.

In our equilibrium study we consider the effect of softening the
spins, that is allowing them to take continuous instead of discrete
values. 
It is usual in the study of spin models to consider soft-spin versions
of discrete spin models. A well-known example is the Ginzberg-Landau
Hamiltonian which is a continuum version of the discrete Ising
model. Other examples occur in the study of spin glass models. For
instance, the soft spin version of the Sherrington-Kirkpatrick (SK)
\cite{sk} model was studied \cite{sz} in the context of dynamics.
The reason for going to soft-spin versions is that they are
often more amenable
\break{1.4in}
 to theoretical approaches. It is usually expected
that qualitatively the soft and hard spin versions should show similar
behaviour. 

For models with multiple ground states, arising out of frustration,
softness may however change the degeneracy completely, as may be seen
in a three-spin example, or as in the present case as we shall show here.
The effect of spin-softening in systems with competing
interactions has been studied earlier by several authors. Seno and
Yeomans \cite{julia} have looked at the effect of softening spins at
the multiphase 
point of a clock-model. They find, using a perturbative method, that
as a softness parameter is varied the system goes through a series of
different ground states. 
In this work we use a similar perturbative method to prove that the
macroscopic degeneracy of the ground state in the ANNNI model is
lifted  by the smallest amount of softness. We then show explicitly 
how the release of the zero-temperature entropy results in qualitative
differences in the low temperature properties of the system. This is
similar to the recently observed phenomena of entropy release in spin-ice
systems \cite{bss}.
We also construct an effective hard-spin Hamiltonian to describe the low-
temperature properties of the soft-spin model. We have also performed
Monte Carlo simulations on the soft-spin model and verified the
low-temperature predictions of the effective Hamiltonian.

In the second part of the paper we look at the dynamical properties
of the system.
As noted before, the ANNNI model at the multiphase point has a large
number of degenerate ground states. It is, therefore, of interest to look at 
dynamical properties of the system at low temperatures. Here we use
Kawasaki dynamics to evolve the system and consider zero
temperature properties only. Thus two nearest
neighbor spins flip with a rate $\gamma $, provided both magnetization
and energy are conserved. This dynamics has been studied earlier by
Das and Barma \cite{diby}. In this paper we extend their
studies by using the correspondence between $W$-matrices for
stochastic processes and quantum spin chains. 

The correspondence between the stochastic Fokker-Planck operator and
quantum chains has often been exploited to derive dynamical
properties. For instance, the scaling, with system size, of the first
excited state of the quantum Hamiltonian gives the dynamical exponent
of the stochastic process. A well-known example where this
correspondence has been used is in exclusion processes \cite{liget},
which are stochastic models of hard-core diffusing particles. For such
processes, it has been possible to exactly calculate the dynamic exponent
by solving the corresponding quantum model, namely the
Heisenberg model \cite{spohn}. 
The dynamics considered by us is very similar to the
symmetric exclusion process (SEP) but with added restrictions on
allowed moves. To see this we first note that with the present dynamics,
nearest neighbor spins with opposite signs flip, provided that the resulting
configuration satisfies the ground state constraint of no three
successive spins having same signs. Identifying up spins with particles
and down spins with holes we see that the dynamics is equivalent to
hard core particles diffusing on a lattice with the constraint that
there cannot be three successive particles or holes. 
An interesting question is whether these rather strong
constraints make the system different from the SEP. Earlier
numerical work \cite{diby} seems to suggest that the dynamics still
behaves like the SEP. We note that there have been some other recent
studies on exclusion processes with constraints on allowed
configurations \cite{alca}. These 
cases are solvable by the Bethe ansatz and show the same behaviour as
the unconstrained model.

Here we address this question of the effect of the constraints by
studying the quantum Hamiltonian.  
By means of exact numerical diagonalization for finite chains and
through analytic bounds, we have tried to understand the differences
and similarities of the present Hamiltonian with the Heisenberg
Hamiltonian for the SEP. We also discuss the different symmetry
properties of the two 
quantum models. The Heisenberg model has full rotational symmetry and
this has several important implications some of which are of direct
relevance in understanding the original stochastic process. For
example it implies that two-point time correlations in the SEP do not
depend on the number of particles. The present model, on the other hand
is only invariant under rotations in the $XY$ plane.

The rest of the paper is divided into two sections. In section (II),
we consider equilibrium properties of the soft-spin model while in
section (III) we consider the dynamics of the hard-spin model. Section
(IV) contains a summary of our main results and a few concluding
remarks.

\section{ Soft-spin ANNNI model }

We consider the following soft spin version of the ANNNI model:
\bea
H_s= \sum_i J ( 2 s_i s_{i+1} + s_i s_{i+2} ) &+& a~g ( s_i^4/4 -
s_i^2/2 ), \nonumber \\
&&~~~~s_i \in (-\infty,\infty) 
\label{softeq}
\eea
where $a$ is a dimensionless parameter which controls the amount of
softness. In the limit $a \to \infty$ we get the hard-spin model.
We will set $g=1$ since there is no loss of generality in doing so.

Let us first look at the ground states of the soft-spin Hamiltonian given by
Eqn. (\ref{softeq}). To do so we look at the extrema of $H_s$ which
are obtained by setting 
${\partial H}/{\partial s_i} = 0$ for all $i$. This gives:
\bea
2 J (s_{i+1} +s_{i-1}) + J (s_{i+2} +s_{i-2}) +a (s_i^3-s_i) = 0
\label{mini} 
\eea   
Solving this set of coupled nonlinear equations in general is very
difficult. However for small values of the parameter $1/a$ we can
obtain the solutions perturbatively.  
For $a \to \infty $ all configurations, $\{ s_i \}$, with $~s_i= 0,\pm
1$ are solutions. Those with $\{s_i= \pm 1\}$ correspond to the minima.
For finite but large $a$ we try to obtain the solutions
perturbatively with $1/a$ acting as the perturbation parameter.
We denote the unperturbed minima by the set $\{ t_i=\pm 1
\}$. Let us try the following perturbative expansion:  
\bea
s_i = \sum_{n=0}^{\infty} t^{(n)}_i (\frac{1}{a})^n, 
\label{series}
\eea
where the coefficients $t_i^{(n)}$ are independent of $a$ and
$t_i^{(0)} \equiv t_i=\pm 1$ correspond to the 
unperturbed solutions in the limit $a \to \infty$. 
Plugging this into Eqn. (\ref{mini}), we get
\bea
&& J [2 (t_{i+1}+ t_{i-1})+ (t_{i+2}+t_{i-2})]   +
\frac{J}{a} [2 ( t^{(1)}_{i+1}+ t^{(1)}_{i-1})+ \nonumber \\
&& ( t^{(1)}_{i+2}+ t^{(1)}_{i-2})]  + 2 t^{(1)}_i+ \frac{1}{a} (3 t_i
(t^{(1)}_i)^2 + 2 t_i^{(2)}) + O(\frac{1}{a^2}) = 0. \nonumber 
\eea
Equating different powers of $1/a$ to zero we then get:
\bea
&& t^{(1)}_i=\frac{-J}{2}[2 (t_{i+1}+ t_{i-1})+(t_{i+2} +
t_{i-2} )] +O(\frac{1}{a}) \nonumber \\
&& t^{(2)}=\frac{-J}{2 } [ 2( t^{(1)}_{i+1} + t^{(1)}_{i-1}) +
(t^{(1)}_{i+2} + t^{(1)}_{i-2}) ]- \frac{3}{2} t_i {(t^{(1)}_i)}^2
\nonumber 
\eea
and so on. 
Thus we get $2^L$ perturbed minima given by the above perturbation
series. The energies corresponding to
these minima can now be found by putting these solutions into the
expression for energy  in Eqn. (\ref{softeq}). We thus get
\bea
&& E = E_0+E_1+E_2+O(1/a^2),~~~~~~~ \rm{where}~~~~~ \nonumber \\
&& E_0 = \frac{-L a }{4} \nonumber \\
&& E_1 = \sum_i J ( 2 t_i t_{i+1} + t_i t_{i+2} ) \nonumber  \\
&& E_2 = \frac{1}{a} \sum_i [ {(t^{(1)}_i)}^2 + 2 J (t_i
t^{(1)}_{i+1} +t_{i+1} t^{(1)}_i) + J ( t_i t^{(1)}_{i+2} + t_{i+2}
t^{(1)}_i) ] \nonumber \\ 
&& = \frac{-J^2}{ 2 a } \sum_i ( 5 + 4 t_i t_{i+1} + 4 t_i
t_{i+2} + 4 t_i t_{i+3} + t_i t_{i+4} ).
\label{ener} 
\eea
In the above expansion, $E_0$ corresponds to the unperturbed energy,
while $E_1$ and $E_2$ 
represent the corrections resulting from the perturbation. 
In the $a \to \infty$ limit the term $E_1$ causes the energy
levels of the $2^L$ minima to split, with
separation between them $\sim O(J)$. We recognize $E_1$ as the
Hamiltonian for the hard-spin ANNNI model. Thus the lowest energy
level is still $\mu^L$-fold degenerate. The term $E_2$ then causes a
further splitting of the ground states into levels with separation
$\sim O(J^2/{a})$. 

To see whether or not the macroscopic degeneracy of the ground state
survives, we need to consider the interaction Hamiltonian
corresponding to the energy term $E_2$. Since we are interested in the
splitting of the lowest energy level of $E_1$, we only consider the
restricted subspace of spin configurations which are ground states of
$E_1$. In this subspace the Hamiltonian corresponding to $E_2$ can be
rewritten as
\bea
H_2=\frac{-3 L J^2}{2 a } - \frac{J^2}{2 a } \sum_i ( 2 t_i t_{i+2} +
4 t_i t_{i+3} + t_i t_{i+4} ).  
\label{effH}
\eea
Thus all the interactions are ferromagnetic. However the ground state
of $H_2$ is not the state with all spins up, since this does not
belong to the subspace of ground states of $E_1$. To find the ground
state, we write the second term in $H_2$, which we denote by
$h_2$, in the following form (the constant factor $J^2/{(2 a )}$ is
suppressed): 
\bea
&& h_2 = -\sum_i ( 2 t_i t_{i+2} + 4 t_i t_{i+3} + t_i t_{i+4} )
 \nonumber \\    
&& ~~ =  -\sum_{i=(4n+1)} \epsilon (t_i,t_{i+1},t_{i+2},t_{i+3} \mid
 t_{i+4},t_{i+5},t_{i+6},t_{i+7} )   \nonumber \\ 
&& \rm{where}~~\epsilon (t_1,t_2,t_3,t_4 \mid t_5, t_6, t_7, t_8) =
t_1 t_3 + t_2 t_4+ 2 t_3 t_5 + \nonumber \\ 
&& 2 t_4 t_6 +  t_5 t_7 + t_6 t_8 + 2 t_1 t_4 + 4 t_2 t_5 + 4 t_3 t_6
+4 t_4 t_7 + 2   t_5 t_8   \nonumber \\
&& +t_1 t_5  + t_2 t_6 + t_3 t_7 +t_4 t_8 
\eea
and the index $n$ runs from $0$ to $(L/4-1)$ (we take $L$ to be an
integral multiple of $4$).
By enumerating the matrix elements $\epsilon (t_1,t_2,t_3,t_4 \mid
t_5,t_6,t_7,t_8)$ for all allowed spin configurations we find that the
lowest energy configuration is obtained for the periodic sequence
$(\uparrow \uparrow \downarrow \uparrow \uparrow \downarrow ...)$ and
the five other configurations 
obtained by translating and flipping this. Thus we find that the
infinite degeneracy of the ground state is removed and instead we get
a six-fold degenerate ground state. We note that the procedure just
outlined provides a straight forward method of finding the ground
state of any spin Hamiltonian.  
By numerically solving Eqn. (\ref{mini}) for small lattice sizes
$(L=12)$ and finding the minimum energy configurations for $a$ large
enough $(a=50)$, we have verified that the perturbative solutions are quite
accurate. 

The fact that softening of the spins results in removal of the
exponential degeneracy of the ground state means that the finite zero
temperature entropy is released and we expect it to show up in the
behaviour of the low temperature specific heat. This leads to the
soft-spin model having low-temperature properties very different from
the hard-spin version as we shall now see.

We note that the hard spin model is easily solvable by transfer-matrix
methods and one can exactly compute various thermodynamic properties. 
In the soft-spin case the transfer-matrix eigenvalue equation
becomes an integral equation which we have not been able to
solve. Hence we have studied the model by Monte Carlo simulations. We
have used a dynamics which allows three kinds of processes; 

(i) single spin-flip moves, 

(ii) moves in which two nearest neighbor spins are
simultaneously flipped and 

(iii) moves which change the length of a spin.   
   
All three kinds of processes occur with usual Metropolis rates. The
reason for allowing both single and double spin-flips is the
following. We find that in the hard-spin case, equilibration times,
with a single-spin flip dynamics, become very large at low
temperatures. On the other hand, allowing for two-spin flips results
in very fast 
equilibration. This is related to the fact that while the single
spin-flip dynamics at $T=0$ is non-ergodic, including double-flips
makes it ergodic. We expect a similar situation even in the case of
soft-spins and so have included both (i) and (ii). Finally (iii) is
necessary since the spins are now continuous variables and we need to
be able to change their lengths.

In order to compare the properties of the soft-spin model with those
of the hard-spin one, it is necessary to subtract from the soft-spin
free energy a part corresponding to the continuum degrees of
freedom. We thus look at the following free energy:
\bea
F= (-1/\beta) [ \ln Tr e^{-\beta H_s} +L \ln (2) - \ln Tr e^{-\beta
H_g} ]  ,
\label{free} 
\eea
where $H_s$ is as in Eqn. (\ref{softeq}), $H_g= \sum_i a 
(s_i^4/4-s_i^2/2)$, and $Tr$ indicates integration over all spin
variables. We note that 
the above expression for the free energy is equivalent to writing the
partition function in the form
\bea
Z= Tr e^{-\beta H} P( \bar s )   ~~~ \rm{with} \nonumber \\
P(\bar s) = \prod_i \frac{2 e^{-\beta a  (s_i^4/4-s_i^2/2)} }{\int
ds_i e^{-\beta a  (s_i^4/4-s_i^2/2)} }.    
\eea
$H$ being the original hard-spin Hamiltonian and $P(\bar s)$ a
probability distribution over the spin variables. In the limit $a \to
\infty$ this exactly reduces to the hard-spin partition function while
at $T \to \infty$ one gets $Z=2^L$.
>From our simulations we get properties corresponding to the first
part of the free energy in Eqn. (\ref{free}). The second part simply
corresponds to a noninteracting system and its properties can be
easily computed numerically. 

In Fig. \ref{soft} we plot the specific heat data $C(T)$ for both the
soft-spin and hard-spin models. The hard-spin result is exact and
corresponds to infinite system size while the soft-spin data is from
simulations on a chain of length $L=24$. The values of various parameters 
used in the simulation were $a=50$ and $J=1$. The high temperature 
$(T > 1)$ data was obtained by averaging over $10^6$ Monte Carlo steps while 
the low temperature data is over $10^7$ steps. As expected we find 
a second peak in the specific heat at low temperatures. For the hard-spin case
the total area under the curve for $C(T)/T$ is equal to $\ln(2/\mu
)$. The ground 
state entropy, $\ln(\mu)$, which is released when the spins are softened, is
mostly accounted for by the area under the low temperature peak.      

The low temperature properties are quite well reproduced by
the effective Hamiltonian, 
$H_2$, which describes the energy levels in the lowest band.  
The thermodynamic properties of $H_2$  can be exactly calculated by
transfer matrix methods,  
both for finite system sizes and in the infinite size limit. In
Fig. \ref{ltemp} we plot the soft-spin low temperature 
simulation data $C(T)$ for two system sizes and compare them with
results obtained 
from the effective Hamiltonian. We see good agreement between the
two. We also show the infinite system size  $C(T)$ curve obtained
from the Hamiltonian $H_2$. It is interesting to note that the peak
value of the specific heat first increases with system size and then
starts decreasing beyond a certain size.

\section{ Kawasaki Dynamics of the hard-spin ANNNI model at the
multiphase point}

As for the usual exclusion process, the quantum Hamiltonian
corresponding to our process can be easily written and is given by:
\bea
{\mathcal H}= {\cal P} \{ \sum_{k=1}^L -[(\sigma^{+}_k
\sigma^{-}_{k+1} &+& \sigma^{-}_k \sigma^{+}_{k+1}) + \nonumber \\
&&  \frac{1}{2}
(\sigma^{z}_k \sigma^{z}_{k+1} -1)] P_k \} {\cal P}  
\label{fkop}
\eea
where $\sigma_k^{\alpha}$ are the usual Pauli matrices, $P_k$ are local
projection operators given by
\bea
P_k= (1-\sigma_{k-2} \sigma_{k-1})  (1-\sigma_{k+2} \sigma_{k+3}) /4
\eea
and ${\cal P}=\prod_{k=1}^{L} P_k$ is a global projection operator
which projects onto the space of 
allowed states, {\it i.e} those that satisfy the ground state
constraint. The spin-flip rate, $\gamma$, has been set to unity.  
Alternatively we can write the Fokker-Planck operator in the
following form:
\bea
{\mathcal H}= -\sum_{k=1}^L ( \theta_k + \theta_k^2 ) \nonumber
~~~~~~{\rm where} \\ 
\theta_k= {\cal P} (\sigma^+_k \sigma^-_{k+1} +\sigma^-_k
\sigma^+_{k+1}) {\cal P}.
\label{fkop2}
\eea 
The term $\sum_k \theta_k^2$ is the diagonal term since it corresponds
to flipping an unequal pair twice.
It is important to write the diagonal part carefully. For
instance if in Eqn. (\ref{fkop}), the local projection operators, $P_k$,
were not present, the off-diagonal elements of ${\mathcal H}$ would still be
correct but the diagonal ones would be wrong.

We now study the properties of this quantum Hamiltonian. Our interests
are (a) to compare the symmetry properties and conservation laws
of the present Hamiltonian with that of the Heisenberg model and
(b) to obtain results on the energy gap and hence the dynamical
exponent. 

\subsection{ Symmetry properties and conservation laws of the quantum
model } 

We first observe that the $z$-component of the total spin, $S^z$,
commutes with ${\mathcal H}$.
This simply implies conservation of spin or number of particles in
the stochastic model. Thus we can classify energy states
into sectors labelled by number of particles $n$. The constraints on
allowed configurations means that for a lattice of length $L$ the
number of particles can vary over the range $[L/3] \leq n \leq
L-[L/3]$ where $[L/3]$ denotes the smallest integer greater than or
equal to $L/3$. It can be shown that except in the lowest and highest
sectors, in every other case the dynamics is ergodic. It then follows
from detailed balance that the steady state is one in which all
allowed configurations in a given sector occur with equal
probability. For the quantum model this means that the ground state in
any sector is an equally weighted sum over all states ( For the special
case where $L$ is a multiple of $3$, the lowest and highest sectors
have $3$-fold degenerate ground states ).      

The other components of the total angular momentum $S^x$ and $S^y$
however do 
not commute with ${\mathcal H}$. Thus the present Hamiltonian has 
$U(1)$ symmetry instead of the $SU(2)$ symmetry of the Heisenberg
model. Also even though the ground states are degenerate, with one
state in every $S^z$ sector, there is no analogue of the
raising/lowering operator  
$S^{\pm}$. If there were such an operator then the entire eigenvalue
spectrum in the $n$-particle sector would be a subset of the
$(n-1)$-particle sector (for $n < L/2$).  
By looking at the spectrum for finite sized lattices we have verified
that this is not so.
 
To study the presence of long-range order in the ground-state, we 
have calculated the two-point static correlation functions
$c_z(r)=<\sigma_0^z  
\sigma_r^z>$ and $c_{\pm}(r)=<\sigma_0^+ \sigma_r^->$ in the ground state
for the half-filled sector. 
The simple characterization of the ground states in
terms of disallowed subsequences enables calculation of
ground-state expectation of any operator by means of transfer matrices.
The transfer matrix method sums over all the different particle
sectors, but in the 
thermodynamic limit the half filled sector dominates, and so we get
correct results (To compute expectation values in other sectors one
would need to introduce a chemical potential).  
Thus we find that
$c_z(r)=A\cos(\phi-2 \pi r/3) e^{-r/\xi}$ where
$\xi=1/\log((3+\sqrt(5))/2)=1.03904...$ and $A$ and $\phi$ are
constants that have different values on odd and even sites.
Fourier transforming $c(r)$ gives the structure factor
$<\sigma^z(-q) \sigma^z(q)>$ which has the form 
shown in Fig. \ref{sq}. We note that it is non-vanishing at all $q$.
The off-diagonal correlation can similarly be obtained using transfer
matrices but the calculation becomes extremely cumbersome. Instead we
have computed this correlation numerically for finite lattices and find
that it saturates, for large $r$, to a constant value, which is given by
$<\sigma_{-}>^2=0.02917...$ (which has been obtained by using the
transfer matrix method). 

Thus we find that ground-state correlation functions show the same behaviour as
in the Heisenberg chain. For the Heisenberg model $c_z(r)$ is
delta-correlated while $c_{\pm}(r)$ saturates to the value $1/4$
(which is much larger than  its value in the present model).
The presence of off-diagonal long range order means that $U(1)$
symmetry is broken in the ground state. This is analogous to the
breaking of $SU(2)$ symmetry in the ground state of the Heisenberg model.
On the other hand, consider the $XXZ$ chain \cite{yang} defined by
the Hamiltonian
\bea
{\mathcal H}=\sum_{k=1}^L -(\sigma^{+}_k \sigma^{-}_{k+1} + \sigma^{-}_k
\sigma^{+}_{k+1}) - \frac{\Delta}{2} \sigma^{z}_k \sigma^{z}_{k+1}.
\eea
Away from the two isotropic points ($\Delta = \pm 1$), this has the
same symmetry as the present 
model. It has no long range order in the gapless phase ($-1 < \Delta <
1$) and all correlations $<\sigma^{\alpha}(0) \sigma^{\alpha}(r)>$ have
power law decays. In the ferromagnetic phase ($\Delta > 1$), the model
has a gap and full ferrromagnetic long-range order in the ground state,
with ultra-local longitudinal correlations namely $c_z(r)=1/4$.  
Thus we see that as far as ground state correlations are concerned the
present model is different from the anisotropic $XXZ$ chain even
though they have the same symmetry properties. Our model is more
similar in properties to the ferromagnet ($\Delta=1$) but has a
nontrivial depletion of the condensate, as well as a nontrivial
$<\sigma^z_0 \sigma^z_r>$ correlation.

Finally we note that rotational invariance of the Heisenberg model means
that two-point time correlations are completely determined by single 
magnon excitations and so have the same behaviour in any $S_z$
sector \cite{spohn,stinch}. This result does not hold in the case of the
present model.

A second conserved quantity in the model is the total linear
momentum. This follows from the translation invariance of ${\mathcal
H}$. The momentum operator 
commutes both with ${\mathcal H}$ and $S^z$ so that in each $S^z$ sector energy
states can be labelled by their momentum. Clearly the ground state has
zero momentum. 

\subsection{ Results on the energy gap}

As is well known the first excited state of ${\mathcal H}$ determines the decay
of correlations for the stochastic process. Thus the energy gap
$\Delta \sim 1/L^z $ and this determines the dynamic exponent $z$.
For the SEP, which corresponds to the Heisenberg ferromagnet, it is
known that $z=2$. This simply reflects the diffusive modes in the
dynamics. The dynamics studied here is very similar to the SEP but
with the constraints on the allowed number of succesive particles and
holes. An interesting question is whether these rather strong
constraints change the dynamical exponent. 
Unlike the Heisenberg model where the Bethe ansatz is applicable and yields
information on the eigenvalue spectrum, the Hamiltonian in 
Eqn. (\ref{fkop}) is much more complicated and we have not been able to
use the Bethe ansatz. We have looked at the eigenvalue spectrum by
numerical diagonalization of ${\mathcal H}$ for small system sizes and also
through Monte Carlo simulations. We also obtain various analytic
bounds on the energy levels. 

\vspace{0.5cm}
{(i)\it Results of numerical diagonalization of ${\mathcal H}$ and
Monte Carlo simulations} 
\vspace{0.5cm}

We have carried out exact diagonalization of the Hamiltonian in
Eqn. (\ref{fkop}) for chains of length upto
$L=22$ at half filling. The diagonalization has been done in the
momentum basis. This makes the Hamiltonian block diagonal and enables
us to go to quite large chian sizes. We find that for small $L$ the first
excited state occurs 
at total linear momentum $q=\pi$ and the gap seems to
decreases as $\sim 1/L$. However from $L=22$ onwards, the first
excited state shifts to $q=2 \pi/L$ and the gap at this momentum
decreases as $\sim 1/L^2$. In Fig. \ref{diag} we show
the numerically obtained gaps at the two momenta as a function of
system size. We also plot corresponding upper bounds on the gaps (to be
derived in the next section). 

We note here that though it is usually the first excited state that
determines the decay of correlations in the stochastic process, it is
possible to construct correlation functions whose decay is governed by
some other eigenvalue. As an example consider the operator  
$ Q= e^{i \frac{\pi}{L} \sum_k k \sigma^z_k} $. This is the so-called
twist operator, first studied by Lieb, Schultz and Mattis
\cite{lieb}. In this case, 
the decay of the correlation, $<Q(0)Q(t)>$ is determined by the lowest
eigenvalue at momentum $\pi$ since the operator carries momentum $\pi$.
In Fig. \ref{qdec} we show the decay constant as determined from
the correlation decay for different system sizes and compare them with those 
obtained from exact diagonalization. The correlation function is
obtained from Monte Carlo simulations and can also be used for larger
system sizes at which numerical diagonalization becomes too difficult.

\vspace{0.5cm}
{(ii)\it Exact Bounds }
\vspace{0.5cm}

We now find upper bounds on the first excited state. Consider the
sector with states which have $n$ overturned spins. 
The bounds are obtained by constructing trial wave functions orthogonal
to the ground state in each sector. Thus consider the operators 
$\sigma^z(q)=\frac{1}{\sqrt{L}} \sum_k \sigma^z_k e^{i k q}$ and the 
twist operator $ Q$ defined in the previous section. Under
translation these operators transform as
\bea
T \sigma^z(q) T^{\dagger} &=&  \frac{1}{\sqrt{L}} \sum_k \sigma_{k+1} e^{i k q}
=e^{-i q} \sigma^z(q) 
\nonumber \\
T Q T^{\dagger} &=& e^{i \frac{\pi}{L} \sum_k k \sigma^z_{k+1}}=e^{i 2
\pi d} Q \nonumber
\eea
where $d=n/L$ is the filling fraction of particles. If $|0_n>$ is the
ground state in the $n$-particle sector, then
the states $\sigma^z(q)|0_n>$ and $Q|0_n>$ have  momenta $q$ and $2\pi d$
respectively and for $q \neq 0$ are orthogonal to the ground 
state, which has zero momentum. Hence the following expectation values
give us two different upper bounds on the gap:
\bea
(a)~~~~e_z &=& \frac{ <\sigma^z(-q) {\mathcal H} \sigma^z(q)> } 
{<\sigma^z(-q) \sigma^z(q)>}
\label{sze} \\ 
(b)~~~e_Q &=& <Q^{\dagger}{\mathcal H} Q> \label{sQe}
\eea  
where $<...>$ denotes ground state expectations. We now evaluate $(a)$
and $(b)$. We shall henceforth restrict ourselves to the half-filled
sector only, though extensions to other sectors can be done.

(a) To evaluate $e_z$ we
first note that the numerator and denominator in Eqn. (\ref{sze}) can be written
in the following equivalent form:
\bea
<\sigma^z(-q) {\mathcal H} \sigma^z(q)> &=& \frac{1}{2} \sum_l e^{i q l} < [ \sigma_1^z, [{\mathcal H},
\sigma_{l+1}^z] ] > \nonumber \\
<\sigma^z(-q) \sigma^z(q)> &=& \sum_l e^{i q l} <\sigma_1^z \sigma_{l+1}^z> 
\label{sHs}
\eea  
The commutator occurring in the above equation can be evaluated and
gives:
\bea 
[ \sigma_1^z, [&{\mathcal H}&,\sigma_{l+1}^z]] = -4 {\cal P}[(\sigma^+_1 \sigma^-_2+
\sigma^-_1 
\sigma^+_2) p_1 (-\delta_{l,L} +\delta_{l,1})+ \nonumber \\
&& 4(\sigma^+_L \sigma^-_1  
+ \sigma^-_L \sigma^+_1) p_L (\delta_{l,L-1} - \delta_{l,L})] {\cal
P^{\dagger}}  
\eea
Inserting this in Eqn. (\ref{sHs}), and using translational invariance
of the ground state we finally obtain:  
\bea
<\sigma^z(-q) & {\mathcal H}& \sigma^z(q)> \nonumber \\ &=& 4 [1-\cos (q) ] < {\cal P} P_1 (
\sigma_1^{+} \sigma_2^{-} + 
\sigma_1^{-} \sigma_2^{+} ) {\cal P^{\dagger}}>  \nonumber \\
&=& 2 [1-\cos(q)] <{\cal P} P_1 (1-\sigma_1^z \sigma_2^z) {\cal P} >
\label{sHs2}
\eea
where the last step has been obtained using the fact that $<0|{\mathcal H}|0>=0$.
As noted before ground-state expectations of any operator can be
computed using transfer matrices.
The expectation value on the rhs of Eqn. (\ref{sHs2}) is thus found to
have the limiting value $({\rm as}~ L \to \infty)$ 
$<{\cal P} P_1 (1-\sigma_1^z \sigma_2^z) {\cal P}>=8-16/\sqrt{5}$.
The Fourier transform of $c_z(r)$, which gives the structure factor
$<\sigma^z(-q) \sigma^z(q)>$, has already been obtained and was
plotted in Fig. \ref{sq}. We note that it is non-vanishing at all $q$.
Finally, from Eqns. (\ref{sze},\ref{sHs}) we get $e_z$ which is plotted in
Fig. \ref{bound1} along with the exact results from finite size
diagonalization. Putting $q=2 \pi/L$ and putting in all numerical
factors, we get the following result: 
\bea
\Delta \le 19.78 \frac{\pi^2}{L^2}  
\eea

(b) We now obtain the other bound using the twist operator, $Q$. We first
note the following properties of $Q$:
\bea
Q^{\dagger} \sigma^{+}_l \sigma^{-}_{l+1} Q | \{ \sigma \}> &=& 
e^{i \frac{2 \pi}{L}} \sigma^{+}_l \sigma^{-}_{l+1} | \{ \sigma \} >
\nonumber \\
Q^{\dagger} \sigma^{-}_l \sigma^{+}_{l+1} Q | \{ \sigma \}> &=& 
e^{-i \frac{2 \pi}{L}} \sigma^{-}_l \sigma^{+}_{l+1} | \{ \sigma \} >. 
\eea
Using these relations we obtain
\bea
<0| &Q^{\dagger} & {\mathcal H} Q |0>  \nonumber \\
&=& -\sum_k  <0| Q^{\dagger} {\cal P} (\sigma^{+}_k
\sigma^{-}_{k+1} + \sigma^{-}_k \sigma^{+}_{k+1} ) P_k {\cal P}Q |0>
\nonumber \\
&& -  \sum_k \frac{1}{2} <0| Q^{\dagger}{\cal P} (\sigma^{z}_k
\sigma^{z}_{k+1} -1) 
P_k {\cal P} Q |0> \nonumber \\
&=& -\cos(2 \pi/L) \sum_k <0| {\cal P} (\sigma^{+}_k
\sigma^{-}_{k+1} + \sigma^{-}_k \sigma^{+}_{k+1} ) P_k {\cal P} |0>
\nonumber \\
&& - \sum_k \frac{1}{2} <0| {\cal P} (\sigma^{z}_k \sigma^{z}_{k+1} -1)
P_k {\cal P} |0>  \nonumber \\
&=& \frac{L}{2} [1-\cos(2 \pi/L)] <0| {\cal P} (1-\sigma^{z}_k
\sigma^{z}_{k+1}) P_k {\cal P} |0>,
\eea
where in the last step we have again used $<O|{\mathcal H}|0>=0$ and translational
invariance of the ground state. The expectation value above has
already obtained so that we get, for large $L$, the
following bound for the gap at momentum $q=\pi$. 
\bea
\Delta \le 0.845 \frac{\pi^2}{L}
\eea
In Fig. \ref{diag} we have plotted both the bounds and the exact finite size
results at $q=2 \pi/L$ and $q=\pi$ as functions of the system size.

\section{Summary}  

In summary we have studied a one-dimensional spin model with competing 
interactions and studied its low-temperature equilibrium and dynamical
properties. In the equilibrium case we have shown that low temperature
properties of the soft-spin and  
hard-spin versions of the model can be very different. The hard-spin 
version of the model has an infinitely degenerate 
ground-state. Through a perturbative calculation we have shown, that as 
soon as we introduce the slightest amount of softness, the degeneracy is 
lifted. The ground state energy levels split to form a band which is 
separated from 
higher levels by $\Delta E =O(J)$. The energy levels within this lowest band
are described by an effective hard-spin Hamiltonian, containing ferromagnetic
interactions upto fourth neighbour terms. This can be used to approximately 
derive the low temperature properties of the model. We find reasonably
good agreement 
with results from Monte Carlo simulations of the soft-spin model.   

Our results indicate that the fixed-length ($a \to \infty$) limit is a
singular one in our model at low temperatures. Since the ground state
of the soft-spin model for large but finite $a$ is only six-fold
degenerate, it would order into one of these six states as $T \to 0$.
This implies the occurrence of a zero-temperature phase transition and
the existence of an appropriately defined correlation length that
diverges as $T$ goes to zero. In the fixed-length ($a \to
\infty$) limit, on the other hand, averaging over all the degenerate
ground states leads to a finite correlation length even at $T=0$. These
results suggest that it would be interesting to study the effects of
softening the spins on the thermodynamic behavior of two- and
higher-dimensional hard-spin models with extensive ground-state
entropy. A well-known model of this kind is the nearest-neighbor Ising
antiferromagnet on a triangular lattice~\cite{triafm}. This model does
not exhibit any phase transition at a non-zero temperature. The
degeneracy-lifting effect of introducing magnitude fluctuations found
in our study suggests that soft-spin versions of this and other similar
models may exhibit finite-temperature phase transitions. Further
investigation of this question would be very interesting.

We believe that the removal of the  exponential ground-state
degeneracy by the introduction of spin-softness in the model studied
here is a special case of a more general phenomenon in which
the presence of additional degrees of freedom allows the system to
relieve frustration and thus reduce the number of degenerate ground
states. Coupling the hard spins to other degrees of freedom, such as
elastic variables describing possible deformations of the underlying
lattice, would probably have similar effects on the degeneracy of the
ground state. It is interesting to note in this context that a 
``deformable'' Ising antiferromagnet on a triangular
lattice in which the Ising spins are coupled to elastic degrees of 
freedom exhibits~\cite{defafm} a Peierls-type phase transition at a 
non-zero temperature. The ordering of the spins at this transition is 
accompanied by a distortion of the lattice. In general, it is expected
that in real, physical systems, such couplings to other degrees of freedom,
however weak, would induce some kind of ordering of the spins as the
temperature is reduced toward zero, thereby avoiding the unstable
situation of having a non-vanishing entropy per spin at $T=0$.

Many disordered spin systems, such as spin glasses~\cite{book}, exhibit
a large number of nearly-degenerate metastable states arising out of
frustration. To take an example, the SK model~\cite{sk} of
infinite-range Ising spin glass is known~\cite{book} to have an
exponentially large number of local minima of the free energy (locally
stable solutions of the TAP equations~\cite{tap}) at sufficiently low
temperatures. These local minima of the free energy become local minima
of the energy at $T=0$. The presence of a large number (divergent in the
thermodynamic limit) of nearly-degenerate metastable states is crucial
in the development of the present understanding~\cite{book} of the
equilibrium and dynamic properties of this system at low temperatures.
Our results about the lifting of degeneracy by the introduction of
spin-softness raise the following interesting question: would the
low-temperature properties of a soft-spin version of the SK model
differ in any significant way from those of the original model? While
soft-spin versions of the SK model have been used in studies~\cite{sz}
of the dynamics, questions about how the number and properties of the
metastable states of this model change as the spins are made soft have
not been addressed in detail. Further investigation of these issues
would be most interesting. 

Finally it is interesting to note that a similar way of lowering
frustration is to make the coupling constants soft while keeping the
spins hard. For example, in the case of the Edwards-Anderson Ising spin
glass model, two versions have been studied~\cite{binder}. One is the
$\pm J$ model where the nearest-neighbor coupling constants randomly
take the discrete values $\pm J$ with equal probability. In the other
case, the $J$s are chosen from a gaussian distribution. In $d=2$, both
these cases are believed to have zero-temperature phase transitions,
but the nature of the transition is different in the two cases. This
difference again arises because of the different ground-state
degenaracies in the two cases. In the $\pm J$ model, the ground-state
is exponentially degenerate, while it is unique (modulo a global
inversion of all the spins) in the gaussian case. However in higher
dimensions where the transition temperature is finite, critical
properties near the transition appear to be the same in both cases.

In our nonequilibrium studies we considered the Kawasaki dynamics
and studied the quantum Hamiltonian corresponding to 
the Fokker-Planck operator for the stochastic process. The spectrum of
the Hamiltonian is obtained by numerical 
diagonalization of finite chains. An interesting crossover of the first
excited state from momentum $\pi$ to $2\pi/L$ is observed with
increase in system size. We have found analytic upper bounds on the
gaps at these two momenta. These, along with our numerical diagonalization
results suggest that the gap vanishes as $\sim 1/L^2$ and so the dynamics 
is diffusive as in SEP. We have also compared the symmetry properties
of our Hamiltonian with the Heisenberg model. We find that while the
model has the symmetry of the $XXZ$ model, its ground-state properties
are closer to those of the ferromagnetic isotropic point. In summary
we have shown that our model is a very nontrivial cousin of the
Heisenberg ferromagnet. The exclusion of three adjacent like spins
essentially changes the model dynamics, and results in a nontrivially    
depleted condensate in $<\sigma^x_0 \sigma^x_r>$ and a nontrivial
gapped $<\sigma^z_0 \sigma^z_r > $ correlation function. The existence
of a groundstate in every $S^z$ sector is quite obvious from the
stochastic point of view but a nontrivial one within the framework of
the quantum system (e.g. the absence of a $\sigma^{-}$ operator), and
require a deeper understanding. 

\section{Acknowledgements} 
We thank Mustansir Barma, Dibyendu Das and Deepak Dhar for useful
discussions. AD is grateful to the Poornaprajna Institute for 
financial support.

\vspace{2cm}

\vbox{
\vspace{0.25cm}
\epsfxsize=8.0cm
\epsfysize=7.0cm
\epsffile{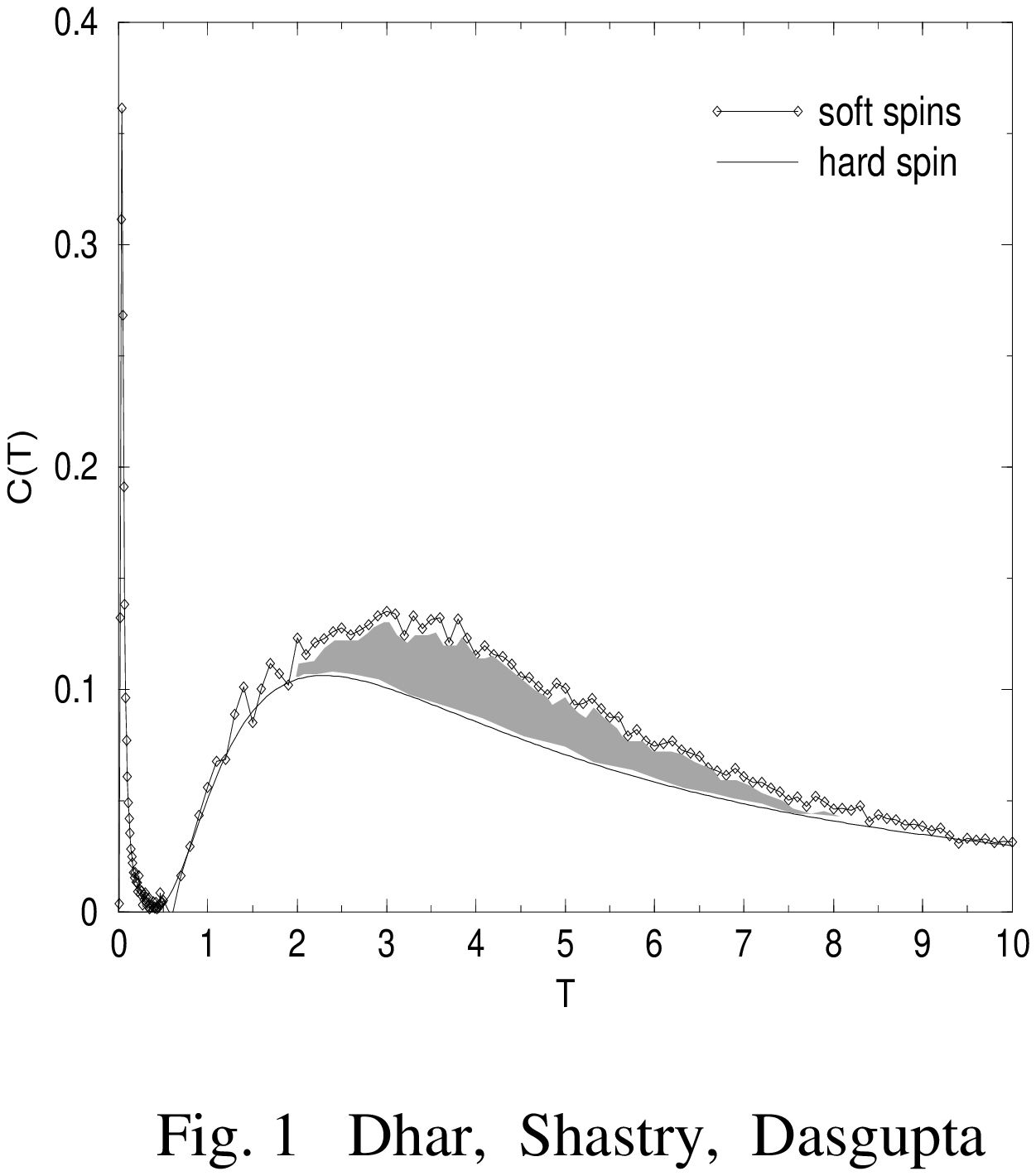}
\begin{figure}
\caption{\label{soft} Simulation data $C(T)$ for the soft-spin model on 
a lattice of size $N=24$. A low temperature peak can be seen. For comparision 
we have also plotted the hard-spin results. 
Most of the entropy released ($\sim 85 \%$) is contained within the
low temperature peak while the remaining occurs in the
high temperature region (shaded portion).  
}  
\end{figure}}

\vbox{
\vspace{0.25cm}
\epsfxsize=8.0cm
\epsfysize=7.0cm
\epsffile{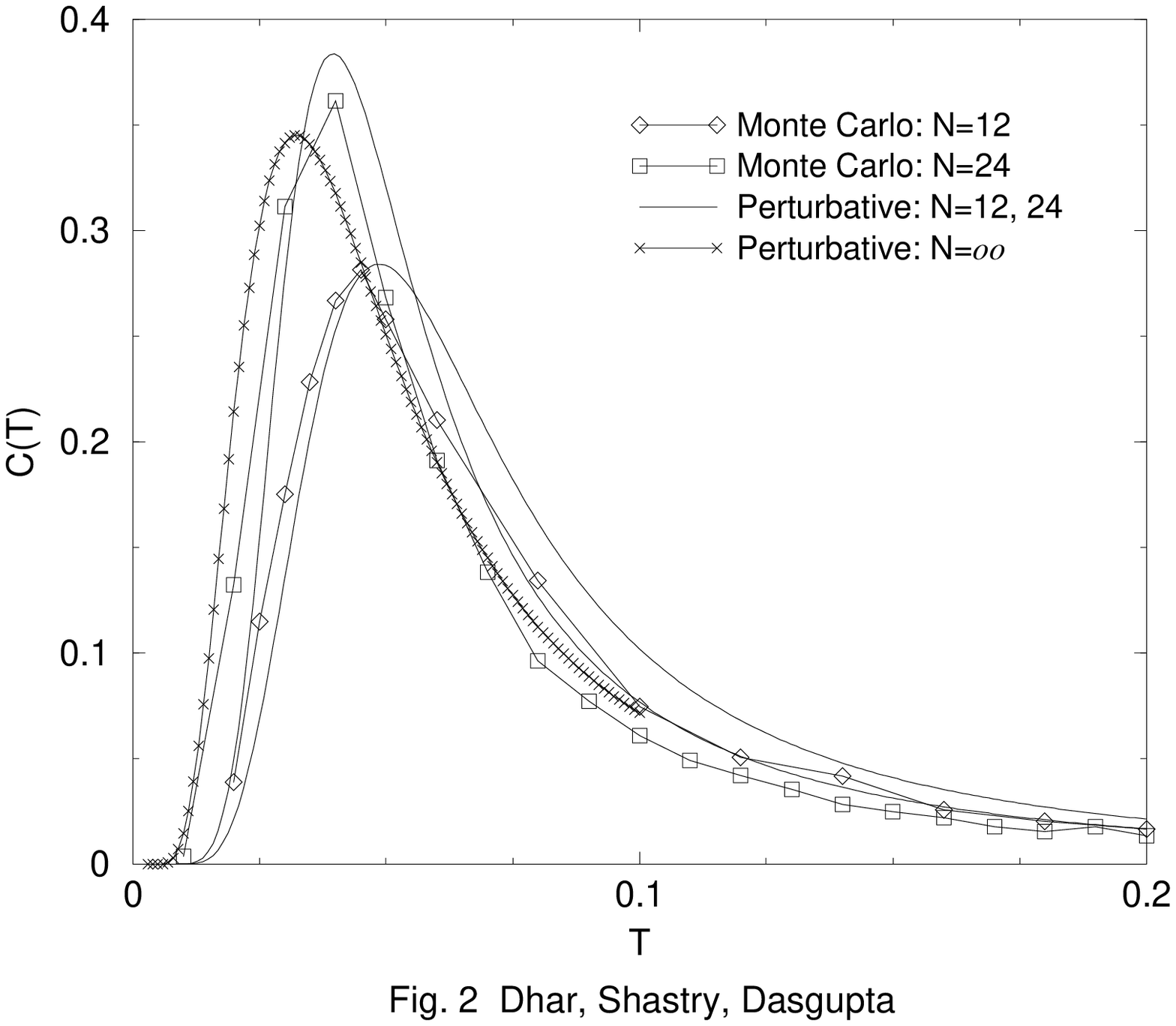}
\begin{figure}
\caption{ \label{ltemp} The plot of $C(T)$ at low temperatures as obtained 
from simulations and from the effective Hamiltonian for different system 
sizes. We also show the effective Hamiltonian result for infinite system 
size.
}  
\end{figure}}

\vbox{
\vspace{0.25cm}
\epsfxsize=8.0cm
\epsfysize=7.0cm
\epsffile{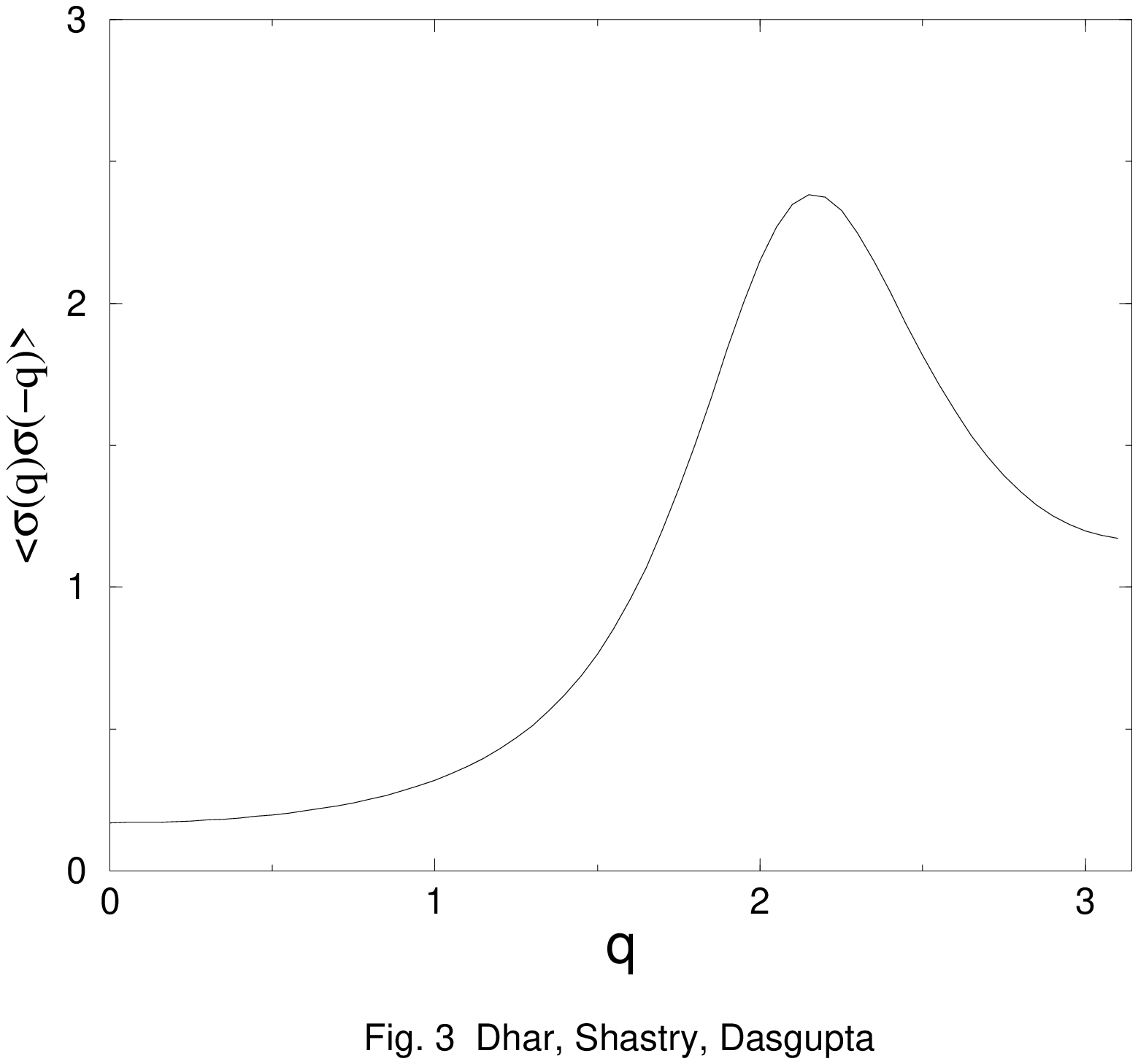}
\begin{figure}
\caption{ \label{sq} 
The diagonal structure factor plotted as a function of the total wave
number $q$.
}  
\end{figure}}

\vbox{
\vspace{0.25cm}
\epsfxsize=8.0cm
\epsfysize=7.0cm
\epsffile{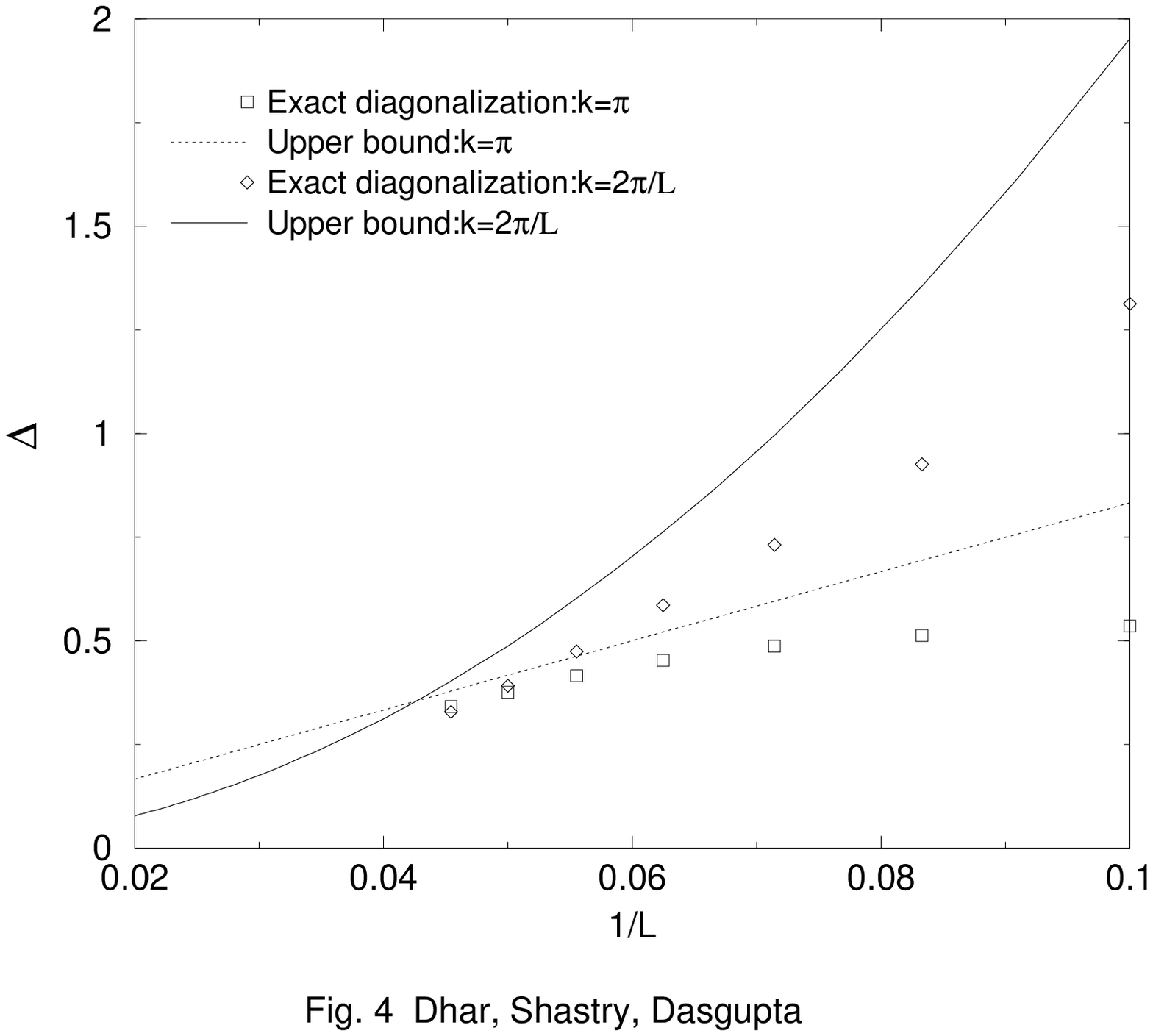}
\begin{figure}
\caption{ \label{diag} 
In this figure the exact energy gaps $\Delta$ at the two momenta $\pi$ and
$2\pi /L$ are plotted against inverse system size. Also plotted are
exact bounds at the two momenta.
}  
\end{figure}}

\vbox{
\vspace{0.25cm}
\epsfxsize=8.0cm
\epsfysize=7.0cm
\epsffile{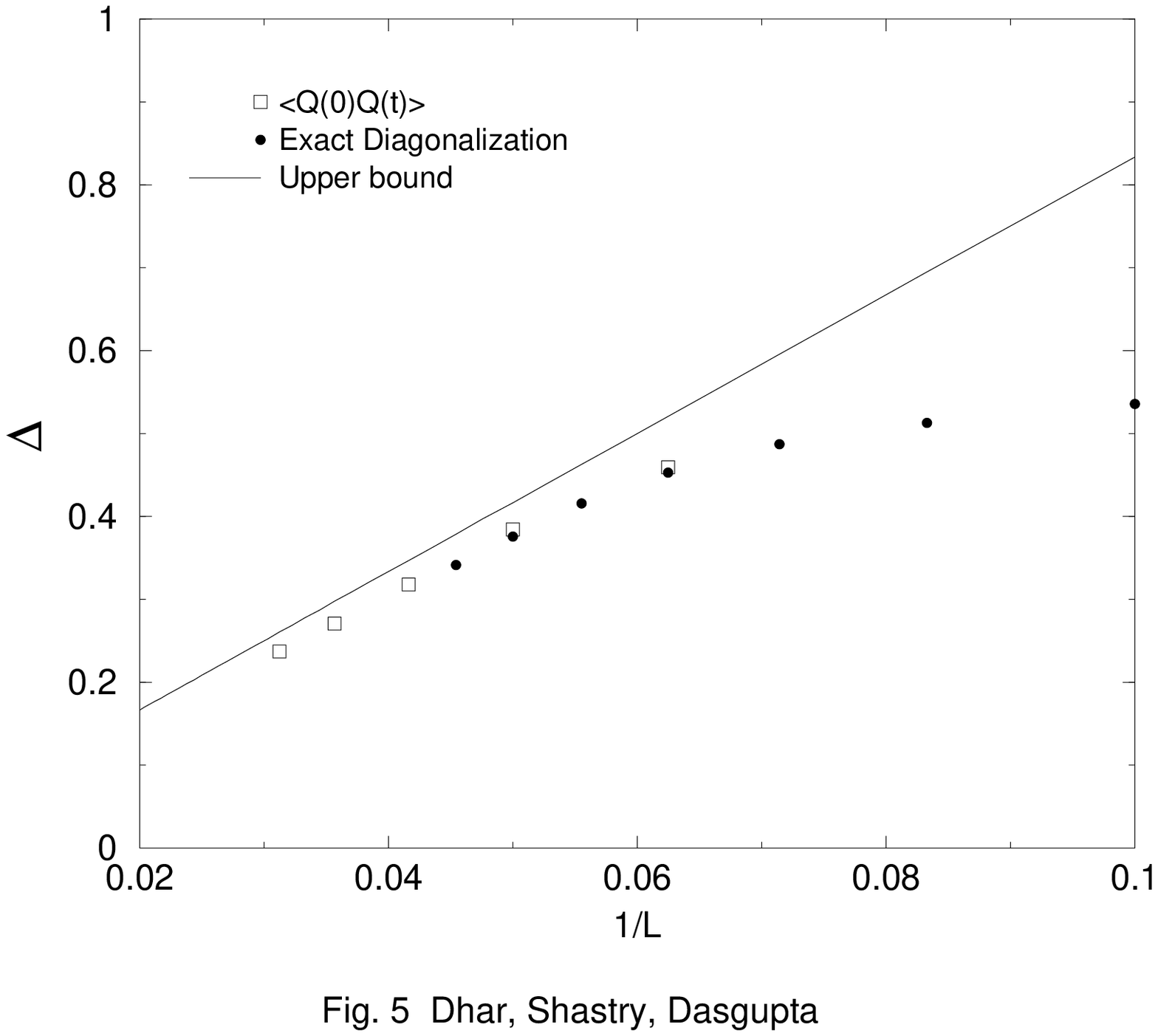}
\begin{figure}
\caption{ \label{qdec} 
The energy gap, as obtained from the decay of the correlation function
$<Q(0)Q(t)>$ is plotted as a function of inverse system size. Also
plotted are the results from exact numerical diagonalization and the upper
bound. The diagonalization has been done till system size $L=22$ while
the $<Q(0)Q(t)>$ data is from Monte Carlo simulations for system size
upto $L=36$. 
}  
\end{figure}}

\vbox{
\vspace{0.25cm}
\epsfxsize=8.0cm
\epsfysize=7.0cm
\epsffile{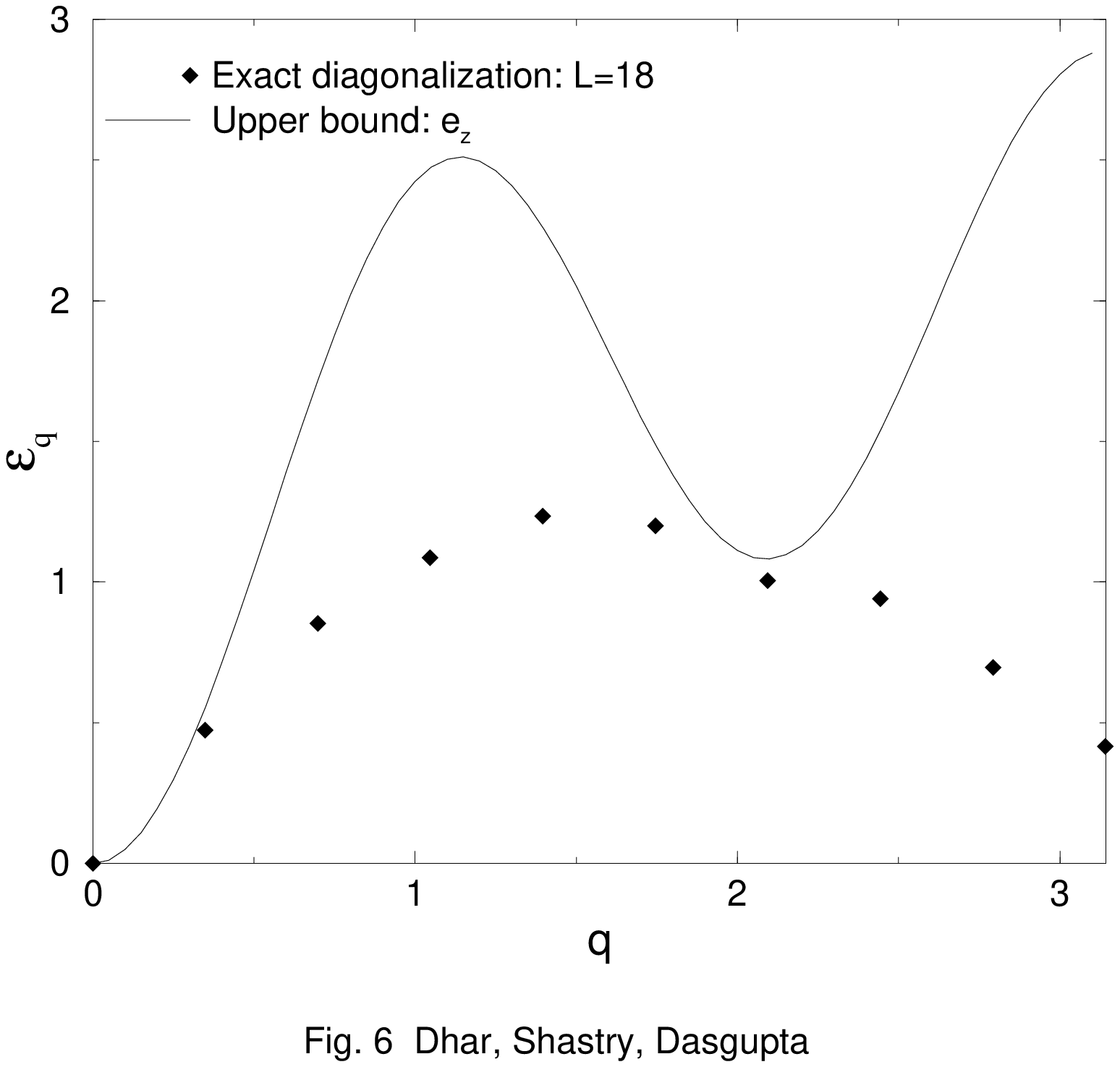}
\begin{figure}
\caption{ \label{bound1} 
The gap upper bound, $e_z$, plotted as a function of total momentum,
$q$. The exact eigenvalues for a system of size $L=18$ are also shown.   
}  
\end{figure}}

\end{document}